# Translational boundaries as incipient ferrielectric domains in antiferroelectric PbZrO$_3$

Ying Liu[1,*], Ranming Niu[2,3], Andrzej Majchrowski[4], Krystian Roleder[5], Julie M. Cairney[2,3], Jordi Arbiol[1,6], Gustau Catalan[1,6,*]

[1]*Catalan Institute of Nanoscience and Nanotechnology (ICN2), Campus Universitat Autonoma de Barcelona, Bellaterra 08193, Spain*

[2]*School of Aerospace, Mechanical and Mechatronic Engineering, The University of Sydney, Sydney, N.S.W. 2006, Australia*

[3]*Australian Centre for Microscopy and Microanalysis, The University of Sydney, Sydney, N.S.W., 2006, Australia*

[4]*Institute of Applied Physics, Military University of Technology, ul. Kaliskiego 2, 00-908 Warsaw, Poland*

[5]*Institute of Physics, University of Silesia, ul. 75 Pułku Piechoty 1, 41-500 Chorzów, Poland*

[6]*Institut Català de Recerca i Estudis Avançats (ICREA), Barcelona 08010, Catalunya*

*ying.liu@icn2.cat; gustau.catalan@icn2.cat

In the archetypal antiferroelectric PbZrO$_3$, antiparallel electric dipoles cancel each other, resulting in zero spontaneous polarisation at the macroscopic level. Yet in actual hysteresis loops, the cancellation is rarely perfect and some remnant polarization is often observed, suggesting the metastability of polar phases in this material. In this work, using aberration-corrected scanning transmission electron microscopy methods on a PbZrO$_3$ single crystal, we uncover the coexistence of the common antiferroelectric phase and a ferrielectric phase featuring an electric dipole pattern of ↓↑↓. This dipole arrangement, predicted by Aramberri *et al.* (2021) to be the ground state of PbZrO$_3$ at 0K, appears at room temperature in the form of translational boundaries that aggregate to form wider stripe domains of the polar phase embedded within the antiferroelectric matrix.

Antiferroelectrics are materials showing an antiparallel (but switchable) alignment of electric dipoles of equal magnitude, so that in the absence of external voltage, the macroscopic net polarisation is zero [1]. Historically, PbZrO$_3$ (PZO) was the first material proposed to be antiferroelectric [2,3], and is regarded as an archetype. Its electric dipoles arrange in a ↑↑↓↓ fashion (Figure 1a) [2]. For antiferroelectrics, applying a large enough electric field can rearrange the electric dipoles in the same direction, causing an antiferroelectric to ferroelectric transition identifiable by a characteristic double hysteresis loop in the polarisation as a function of the electric field [3–5]. This antipolar-to-polar switching is accompanied by giant charge storage, volume expansion and temperature drop, and hence is promising in applications in high-density capacitors [6–8], high strain transducers [9,10] and electrocaloric cooling [11,12].

Closely related to antiferroelectrics, ferrielectric phases (characterized by possessing antiparallel but uncompensated electric dipoles) have also attracted attention [13–18]. They are reported to exist in different forms and under various conditions. For example, a ↑↑↓↓ dipole pattern was observed in chemically doped PZO [15,19]; in pure PZO, antiparallel electric dipoles with imbalanced magnitude were theoretically predicted to exist under an electric field [13,14], and a more complex ferrielectric structure with a ↑↑↑↑↓↑↑↓ dipolar configuration was also proposed based on a combination of in-situ biasing X-ray diffraction and simulation results [17]. Interestingly, even the ground state of PZO has been proposed to be ferrielectric instead of antiferroelectric, as ab initio calculations by Aramberri *et al.* [16] suggest that a ↓↓↑ dipole pattern could be the lowest-energy state in PZO at 0K and possibly up to room temperature where, being polar, it may contribute to the open double hysteresis loop in PZO [20,21]. Yet, this structure has not been experimentally observed.

In this work, we uncover the existence of the ↓↑↓ ferrielectric phase in PZO single crystal at room temperature. The ferrielectric periodicity is one dipole smaller than the antiferroelectric one, and therefore intercalated ferrielectric layers fulfil the role of translational boundaries (TBs). TBs are discontinuities in the periodic modulation of the antiferroelectric lattice and are intrinsic topological defects in PZO [22–24]. A schematic explanation of such translational boundaries in antiferroelectrics is shown in Figure 1b. The concept can be understood by viewing the antipolar ordering as a square wave modulation of the polarisation, with the translational boundaries (blue dotted lines in Figure 1b) marking a shift in the phase of this square wave by 1/4, 1/2 and 3/4 unit cells, respectively. Domains on either side of a translational boundary are thus related by phase shifts of π/2, π, and 3π/2 [23,24]. The breaking of translational symmetry implies a breaking of the perfect dipole cancellation, and antiphase boundaries in antiferroelectrics are expected to be polar [25]. Wei *et al.* proved the polar nature of antiphase boundaries (translational boundaries with a phase shift of π) in 2014 and highlighted their potential in information storage applications [22].

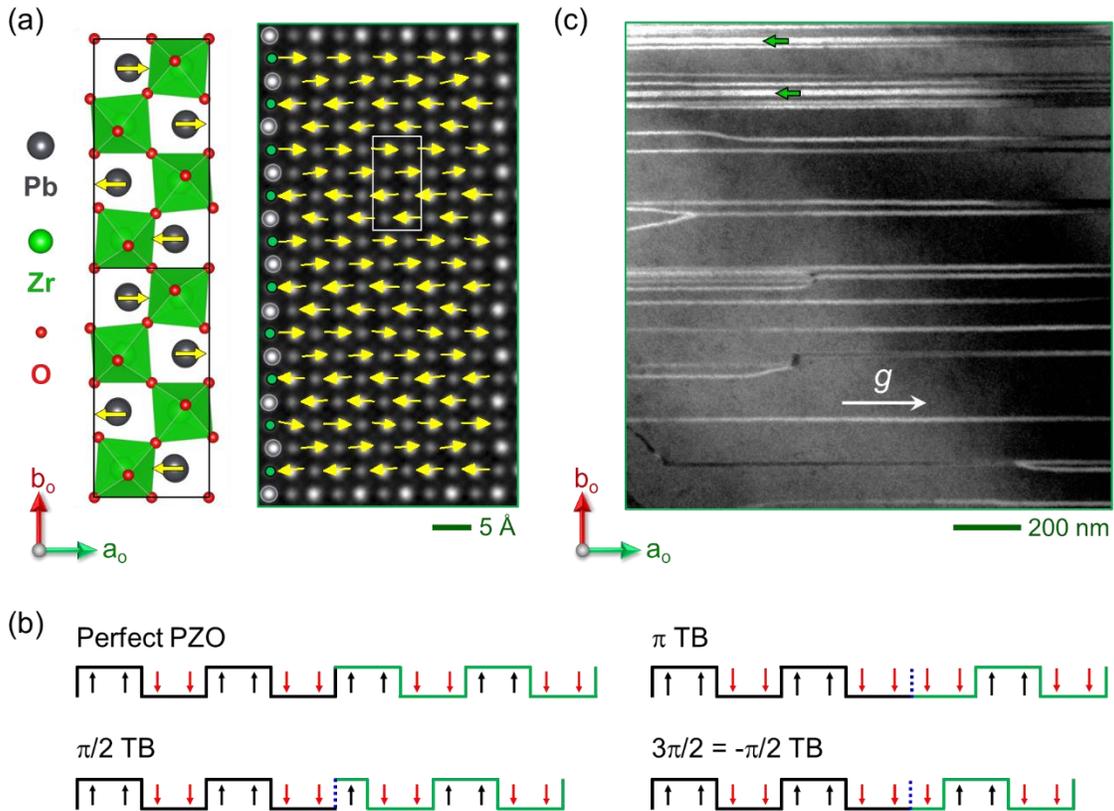

FIG. 1. (a) A structure model of PZO (two unit cells outlined by black rectangles) and a Pb displacement (Pb displacement with respect to their four nearest Zr, yellow arrows) map obtained from a STEM-HAADF image of the PZO single crystal visualized along the $c_O$ axis. (b) Schematics showing electric dipole arrangements and square waves representation of the perfect PZO, PZO with π/2, π, and 3π/2 TBs. Blue dotted lines denote translational boundaries. (c) A transmission electron microscopy bright field image showing TBs observed in PZO single crystal. Green arrows indicate thicker TBs.

Ferrielectric phases and translational boundaries thus share commonalities. Both are closely related to, and often appear within, antiferroelectrics, while at the same time being polar. In this work, we show that, in pure PZO, a ferrielectric phase with the symmetry Ima2 that was predicted to be the ground state of this material, can exist at room temperature in the form of stripe domains that act as translational boundaries of varying thickness and with phase shifts that are always integer multiples of π/2.

A single crystal of PZO was used as the sample for this study. Details of the single crystal fabrication method are provided elsewhere [26]. Electron-transparent lamellae were cut from the crystal using focussed ion beam lithography. The lamellae were heated up to 250 °C and cooled down to favour the nucleation of different domains upon cooling. The samples show the expected antiferroelectric structure of PZO (Figure 1a), but also extended planar structures (Figure 1c). These are similar to those observed by Wei *et al*. [22], who identified them as π translational boundaries, i.e. boundaries that change the phase of the dipole arrangement by a factor of π. We found, however, that the linear structures in the sample have different thicknesses. Green arrows have been added in Figure 1c to mark some thicker ones. Further atomic-scale investigations of these stripe-like features, displayed in Figures 2 – 4, show that they are in fact ferrielectric domains that act as translational boundaries with different phase angles multiple of π/2.

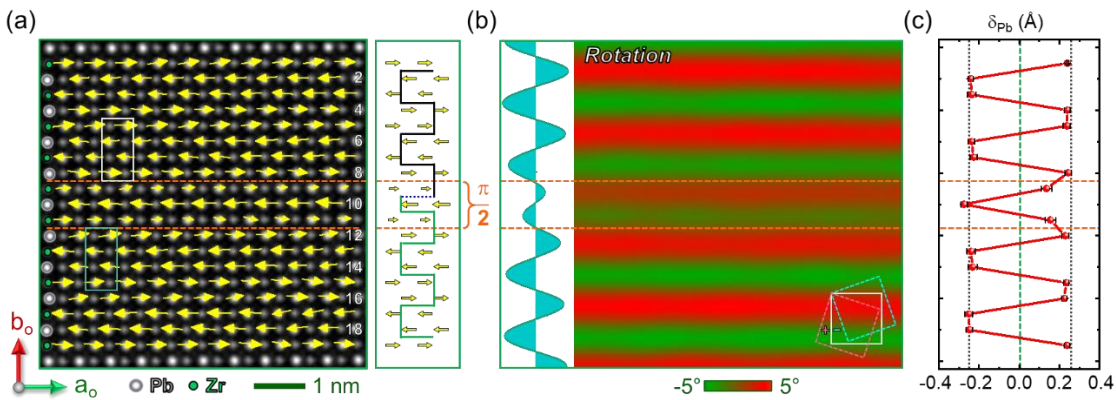

FIG. 2. (a) A Pb displacement map superimposed on the corresponding STEM-HAADF image showing a TB separating regions with π/2 difference in phase (along **b**$_O$ direction). (b) The GPA lattice rotation map of (a). The blue inset curve is an intensity profile of the lattice rotation map. (c) The Pb displacement curve by averaging rows from 1 to 19. Error bars, standard deviation. Black dotted lines, Pb displacement in ideal antiferroelectric PZO.

Figure 2a displays a scanning transmission electron microscopy high angle annular dark field (STEM-HAADF) image of the thinnest translational boundaries found in the PZO single crystal. Pb displacements ($\delta_{Pb}$) with respect to their four nearest Zr were extracted using Python with the "Atomap" library [27]. The obtained $\delta_{Pb}$ map was superimposed on the corresponding STEM-HAADF image. From the $\delta_{Pb}$ map, a disturbance of antiferroelectric order can be observed. Electric dipoles arrange in the ↑↑↓↓↑↓↓↑↑ manner, where an upward dipole is missed, and the single unpaired dipole is larger in magnitude. On both sides of this dipole, the antiferroelectric domains have a phase difference of π/2 and a relative shift of 1/4 orthorhombic unit cells (i.e., one perovskite unit cell) along the orthorhombic **b** direction. We henceforth identify this structure as a π/2 TB.

We have also examined the lattice rotations by means of Geometric Phase Analysis (GPA) [28,29] on the atomic resolution STEM-HAADF images. A GPA lattice rotation map from Figure 2a is shown in Figure 2b. The lattice rotation angle at the TB region and its modulation period is smaller than in other upper and lower regions. The intensity profile indicates this trend. Based on this characteristic in GPA, it is possible to identify the TB even without a $\delta_{Pb}$ map.

Quantitative analysis also shows that the magnitude of the middle unpaired electric dipole of the TB is bigger than the two satellite dipoles on either side, i.e., the dipolar structure is ↓↑↓. This means that, though the internal symmetry of the π/2 translational boundary is polar, its net polarisation can be positive, negative, or zero depending on the relative difference between the central dipole and the sum of the two satellites. Moreover, we find that this relative ratio can change continuously within the same TB. For example, on the far left of the TB of Figure 2a, the middle $\delta_{Pb}$ (0.25 Å) is smaller than the sum of the satellite dipoles (-0.17 Å and -0.21 Å, respectively), while on the far right of the TB, the middle dipole ($\delta_{Pb}$, 0.31 Å) is bigger than the sum of the two satellites ($\delta_{Pb}$, -0.07 Å and -0.15 Å, respectively). In fact, despite the TB unit cell being polar, we find that the average polarisation is close to zero. The averaged $\delta_{Pb}$ plot as a function of the atomic rows is shown in Figure 2c. The middle $\delta_{Pb}$ (0.275 Å) is a little bigger than that in the antiferroelectric region and in the ideal PZO model (black dotted lines), while the satellite $\delta_{Pb}$ (average values of 0.135 Å and 0.155 Å) are much smaller than the middle $\delta_{Pb}$, but the sum of the two (0.135 + 0.155 = 0.290 Å) is almost equal the antiparallel displacement of the central dipole (0.275 Å).

The conclusion from this analysis is that the TB can change its internal polarization from positive to negative or even zero while still preserving the relative sign of the internal displacements, i.e. modulating only their relative magnitude. This is a qualitative difference with respect to ferroelectrics, for which changing the sign of polarization requires inverting the sign of the atomic displacements within the unit cell. The ability to modulate the sign of the polarization without having to overcome a discrete energy barrier means that the internal polarization of the TBs can adapt to local variations in electric fields.

In addition to the π/2 TB, we have also observed TBs with two and three ↓↑↓ dipole units (Figures 3a, 3b). By extrapolating the antiferroelectric domains across these regions, we see that the square waves that represent the antipolar modulation on both sides of the TBs shift by a period difference of 1/2 (Figure 3a) and 3/4 (Figure 3b), respectively, and therefore these are TBs with phase differences of π and 3π/2. Translational boundaries with four or more ↓↑↓ dipole units have never been considered in previous research, yet they are also observed in our experiment. Figures 3c – 3e show the superimposed STEM-HAADF micrographs + $\delta_{Pb}$ maps and GPA lattice rotation maps for even wider TBs. From these images, four (Figure 3c), five (Figure 3d) and seven (Figure 3e) π/2 TB structural units can be determined.

The reason for the formation of π/2, π and 3π/2 TBs is theoretically clear: they are inevitable consequences of the nucleation of antiferroelectric domains at atomic sites separated by a distance that is not an exact multiple of 4 perovskite unit cells. As such, these TBs are protected by symmetry – to eliminate a translational boundary. It would be necessary to rearrange all the dipoles of one of the adjacent domains. In contrast, TBs with a phase difference of 2π (equal to four π/2 TBs), such as in Figure 3c, do not in principle enjoy such topological protection because there is no phase difference between domains separated by TBs with a phase difference of 2π. Put another way: 4 ↓↑↓ dipole units can be replaced by 3 ↓↑↑↓ antiferroelectric dipole units without disturbing the translational symmetry of the adjacent domains. Similarly, five or seven π/2 TB can, in theory, be replaced by one antiferroelectric structural unit and one or three π/2 TB structural units, which would in principle lower the crystal's energy if the antiferroelectric state is the ground state.

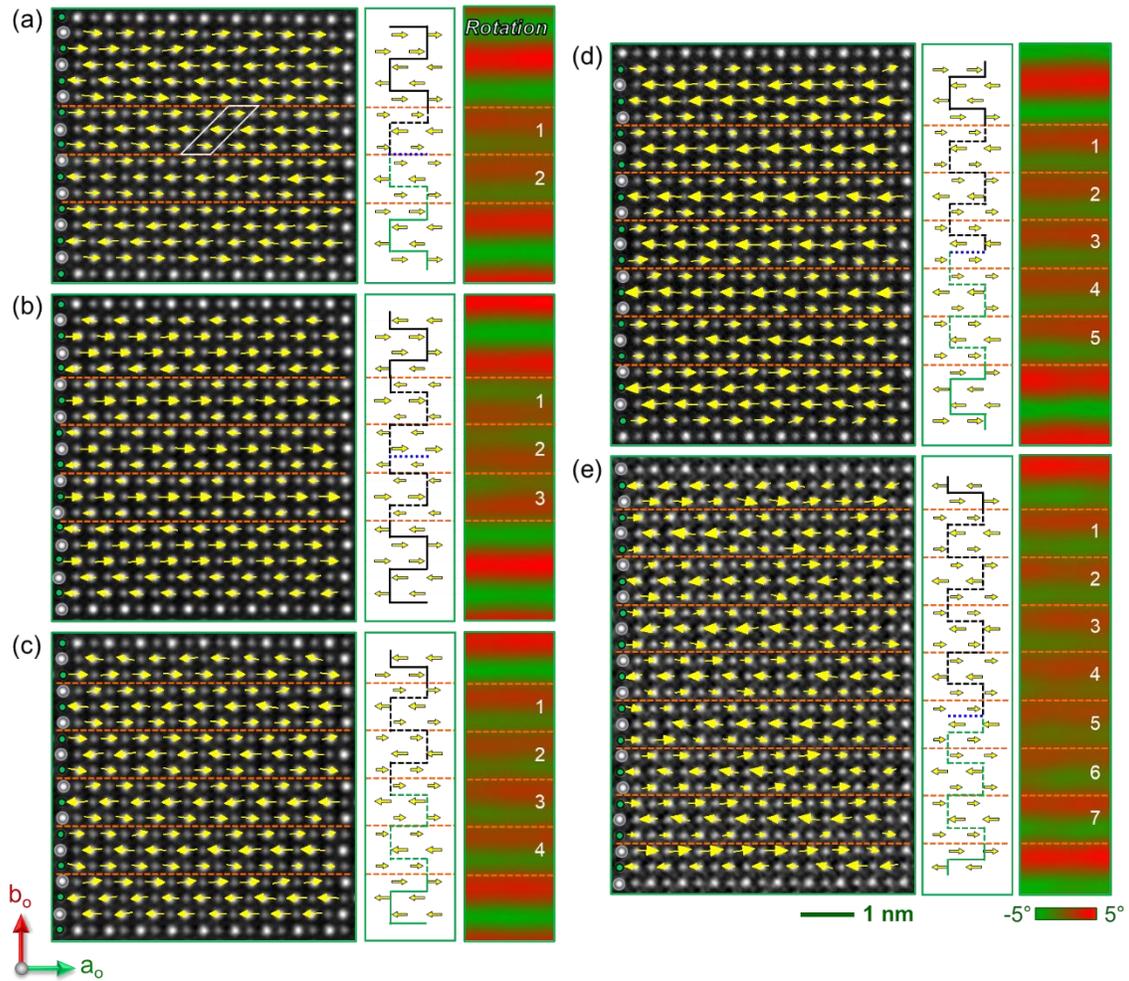

FIG. 3. TBs or ferrielectric phases observed in PbZrO$_3$ single crystal with (a) two, (b) three, (c) four, (d) five and (e) seven ↓↑↓ structural units, corresponding to phase differences of π, 3π/2, 0, π/2 and 3π/2, respectively. A ferrielectric structural unit is outlined by a white parallelogram in (a). A superimposed STEM-HAADF image + Pb map, a Pb displacement schematic and a GPA lattice rotation map are included in each panel. In Pb displacement

schematics, the square waves are extended from outside (solid lines) to inside (dashes lines) of TBs until they meet, showing phase differences of π, 3π/2, 0, π/2, 3π/2, respectively. TBs, blue dotted lines.

At this point, then, it becomes necessary to reexamine whether these structures should be regarded as translational boundaries. They can also be treated as domains of a ferrielectric phase (↓↑↓) embedded within the antiferroelectric matrix (↓↑↑↓). Indeed, the three-dipole arrangement of the TBs is the same as in the unit cell proposed by Aramberri *et al.* [16] for the theoretical ground state of PZO. Our observations suggest that this ferrielectric phase is sufficiently close in energy to the antiferroelectric state, and it can be locally stabilized by translational boundaries at room temperature.

In order to fully characterize the ferrielectric unit cell, it is necessary to determine its oxygen positions. In STEM-HAADF images, only Pb and Zr can be observed. Instead, we turn to STEM integrated differential phase contrast (iDPC) imaging, which is sensitive to light elements [30] and can image oxygens in perovskite oxides [31]. The results for a π/2 and π TB are shown in Figures 4a and 4b, respectively. A primitive cell is outlined using a white parallelogram in Figure 4a. Along the $a_O$ direction, all horizontal oxygen chains are slightly rippled, and tilt in opposite directions on both sides of the biggest Pb displacements (central dipole in the TB structural units). Along the $b_O$ direction, the tilting pattern is / − | −\, that is, clockwise, straight, anticlockwise, with the central (bigger) dipole coinciding with the untilted oxygen chain and showing the characteristic anti-correlation between tilts and polarisations in perovskites. A unit cell was outlined by a white rectangle in Figure 4b, and a possible structural model is shown in Figure 4c. The structure on the x-y plane is supposed to be the same as that reported in Ref. [16], for which the space group is Ima2.

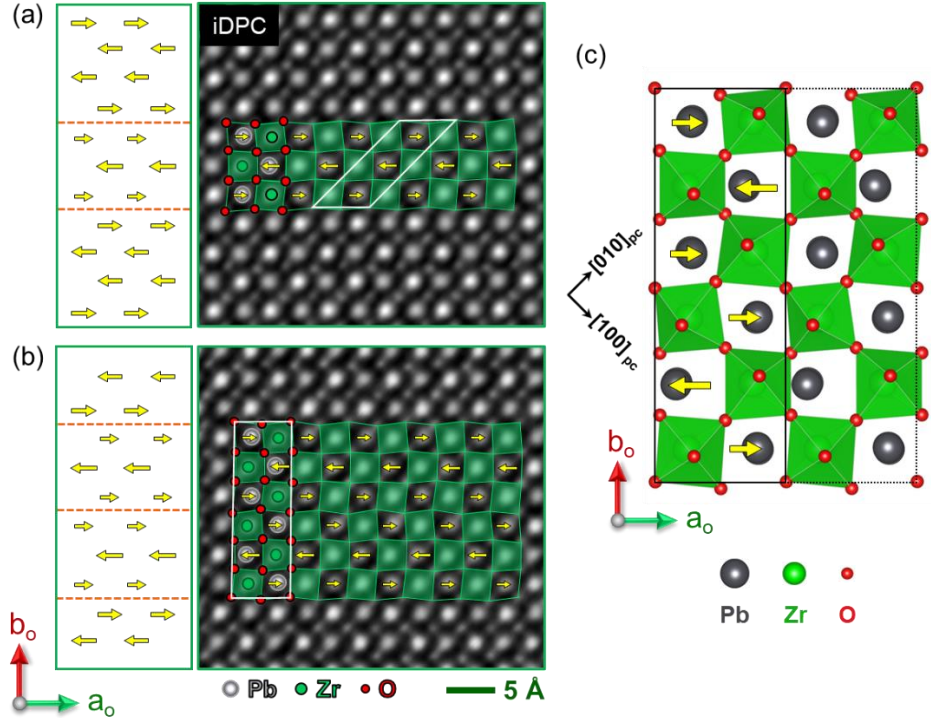

FIG. 4. (a, b) STEM-iDPC images showing Pb, Zr and O distribution at $\pi/2$ and $\pi$ TBs, respectively. (c) The ferrielectric PZO model. Two unit cells are outlined using black rectangles.

In PZO, the antiferroelectric dipole sequence quadruples the size of a single perovskite unit cell, and therefore random nucleation of antiferroelectric domains will inevitably result in TBs. The present work indicates that such TBs can grow or aggregate to form wider stripe domains where the phase difference is larger than $3\pi/2$, the maximum theoretical limit [22]. The internal symmetry of these structures appears to be the same as that predicted in Ref. [16] for the ground state of PZO. Translational boundaries thus act as ferrielectric precursors that may contribute to the small remnant polarisation at 0V often reported in the double hysteresis loops of PZO [20,21,32,33].

**ACKNOWLEDGEMENTS**

This project has received funding from the European Union's Horizon 2020 research and innovation program under grant agreement N° 766726 (TSAR), in addition to Grant PID2019-108573GB-C21 funded by MCIN/AEI/10.13039/501100011033. ICN2 acknowledges funding from Generalitat de Catalunya 2017 SGR 327. ICN2 is supported by the Severo Ochoa program from Spanish MINECO (Grant No. SEV-2017-0706) and is funded by the CERCA Programme / Generalitat de Catalunya. YL acknowledges the BIST Postdoctoral Fellowship Programme (PROBIST) funded by the European Union's Horizon 2020 research and innovation programme under the Marie Sklodowska-Curie grant agreement No. 754510. KC and AM acknowledge funding by the National Science Centre, Poland, grant number 2020/37/B/ST3/02015. The authors are grateful for the scientific and technical support from the Australian Centre for Microscopy and Microanalysis (ACMM) as well as the Microscopy Australia Node at the University of Sydney. Thanks Prof. Jorge Íñiguez and Dr. Hugo Aramberri from the Luxembourg Institute of Science and Technology (LIST), for the discussion of the space group and Glazer notation of the ferrielectric phase.

# REFERENCES

[1] C. Kittel, Phys. Rev. **82**, 729 (1951).

[2] E. Sawaguchi, H. Maniwa, and S. Hoshino, Phys. Rev. **83**, 1078 (1951).

[3] G. Shirane, E. Sawaguchi, and Y. Takagi, Phys. Rev. **84**, 476 (1951).

[4] X. Tan, C. Ma, J. Frederick, S. Beckman, and K. G. Webber, J. Am. Ceram. Soc. **94**, 4091 (2011).

[5] P. Vales-Castro, M. Vellvehi, X. Perpiñà, J. M. Caicedo, X. Jordà, R. Faye, K. Roleder, D. Kajewski, A. Perez-Tomas, E. Defay, and G. Catalan, Adv. Electron. Mater. **7**, 2100380 (2021).

[6] A. Chauhan, S. Patel, R. Vaish, and C. R. Bowen, Materials **8**, 8009 (2015).

[7] Z. Liu, T. Lu, J. Ye, G. Wang, X. Dong, R. Withers, and Y. Liu, Adv. Mater. Technol. **3**, 1800111 (2018).

[8] C. A. Randall, Z. Fan, I. Reaney, L.-Q. Chen, and S. Trolier-McKinstry, J. Am. Ceram. Soc. **104**, 3775 (2021).

[9] W. Y. Pan, C. Q. Dam, Q. M. Zhang, and L. E. Cross, J. Appl. Phys. **66**, 6014 (1989).

[10] S.-E. Park, M.-J. Pan, K. Markowski, S. Yoshikawa, and L. E. Cross, J. Appl. Phys. **82**, 1798 (1997).

[11] R. Pirc, B. Rožič, J. Koruza, B. Malič, and Z. Kutnjak, EPL (Europhysics Letters) **107**, 17002 (2014).

[12] P. Vales-Castro, R. Faye, M. Vellvehi, Y. Nouchokgwe, X. Perpiñà, J. M. Caicedo, X. Jordà, K. Roleder, D. Kajewski, A. Perez-Tomas, E. Defay, and G. Catalan, Phys. Rev. B **103**, 054112 (2021).

[13] P. Tolédano, and M. Guennou, Phys. Rev. B **94**, 014107 (2016).

[14] P. Tolédano, and D. D. Khalyavin, Phys. Rev. B **99**, 024105 (2019).

[15] Z. Fu, X. Chen, Z. Li, T. Hu, L. Zhang, P. Lu, S. Zhang, G. Wang, X. Dong, and F. Xu, Nat. Commun. **11**, 3809 (2020).

[16] H. Aramberri, C. Cazorla, M. Stengel, and J. Íñiguez, Npj Comput. Mater. **7**, 196 (2021).

[17] R. G. Burkovsky, G. A. Lityagin, A. E. Ganzha, A. F. Vakulenko, R. Gao, A. Dasgupta, B. Xu, A. V. Filimonov, and L. W. Martin, Phys. Rev. B **105**, 125409 (2022).

[18] L. Qiao, C. Song, Q. Wang, Y. Zhou, and F. Pan, ACS Appl. Nano Mater. **5**, 6083 (2022).

[19] T. Ma, Z. Fan, B. Xu, T. H. Kim, P. Lu, L. Bellaiche, M. J. Kramer, X. Tan, and L. Zhou, Phys. Rev. Lett. **123**, 217602 (2019).

[20] L. Pintilie, K. Boldyreva, M. Alexe, and D. Hesse, J. Appl. Phys. **103**, 024101 (2008).

[21] M. Guo, M. Wu, W. Gao, B. Suna, and X. Lou, J. Mater. Chem. C **7**, 617 (2019).

[22] X. K. Wei, A. K. Tagantsev, A. Kvasov, K. Roleder, C. L. Jia, and N. Setter, Nat. Commun. **5**, 3031 (2014).

[23] I. Rychetsky, W. Schranz, and A. Troster, Phys. Rev. B **104**, 224107 (2021).

[24] X. K. Wei, C. L. Jia, K. Roleder, and N. Setter, Mater. Res. Bull. **62**, 101 (2015).

[25] G. Catalan, J. Seidel, R. Ramesh, and J. F. Scott, Rev. Mod. Phys. **84**, 119 (2012).

[26] J.-H. Ko, M. Górny, A. Majchrowski, K. Roleder, and A. Bussmann-Holder, Phys. Rev. B **87**, 184110 (2013).

[27] M. Nord, P. E. Vullum, I. MacLaren, T. Tybell, and R. Holmestad, Adv. Struct. Chem. Imaging **3**, 9 (2017).

[28] M. J. Hytch, E. Snoeck, and R. Kilaas, Ultramicroscopy **74**, 131 (1998).

[29] Y. Liu, Y. J. Wang, Y. L. Zhu, C. H. Lei, Y. L. Tang, S. Li, S. R. Zhang, J. Li, and X. L. Ma, Nano Lett. **17** , 7258 (2017).

[30] I. Lazić, E. G. T. Bosch, and S. Lazar, Ultramicroscopy **160**, 265 (2016).

[31] Y. Liu, R.-M. Niu, S. D. Moss, P. Finkel, X. Z. Liao, and J. M. Cairney, J. Appl. Phys. **129**, 234101 (2021).

[32] J. Ge, D. Remiens, X. Dong, Y. Chen, J. Costecalde, F. Gao, F. Cao, and G. Wang, Appl. Phys. Lett. **105**, 112908 (2014).

[33] K. Boldyreva, D. Bao, G. L. Rhun, L. Pintilie, M. Alexe, and D. Hesse, J. Appl. Phys. **102**, 044111 (2007).